\documentclass{elsart}

\usepackage{graphicx}              

\begin{document}

\begin{frontmatter}

\title{Heavy Residues with A$<$90 from the Asymmetric Reaction 
of 20 AMeV $^{124}$Sn+$^{27}$Al as a Sensitive Probe 
of the Onset of Multifragmentation}

\author[iopsas,cyctamu]{M. Veselsky}$^{}$, \thanks{\footnotesize  
             E-mail address: fyzimarv@savba.sk (M. Veselsky).}
\author[cyctamu]{G.A. Souliotis},   
\author[cyctamu]{G. Chubarian},     
\author[cyctamu]{L. Trache},
\author[cyctamu]{A. Keksis},
\author[cyctamu]{E. Martin},
\author[cyctamu]{A. Ruangma},
\author[cyctamu]{E. Winchester}, and
\author[cyctamu]{S. J. Yennello}.

\address[iopsas]{Institute of Physics, Slovak Academy of Sciences, 
           Dubravska 9, 84228 Bratislava, Slovakia }
\address[cyctamu]{Cyclotron Institute,
           Texas A\&M University, College Station, TX 77843, USA }

\begin{abstract}

The cross sections and velocity distributions of heavy residues 
from the reaction of 20 AMeV $^{124}$Sn + $^{27}$Al 
have been measured at forward angles using the MARS recoil separator 
at Texas A\&M in a wide mass range. A consistent overall description 
of the measured cross sections and velocity distributions was achieved 
using a model calculation employing the concept 
of deep-inelastic transfer for the primary stage of peripheral 
collisions, pre-equilibrium emission and incomplete fusion for the primary 
stage of more violent central collisions and the  statistical 
model of multifragmentation (SMM code) for the deexcitation stage. 
An alternative calculation employing the sequential binary decay 
(GEMINI code) could not reproduce the observed yields of the 
residues from violent collisions (A$<$90) due to different kinematic 
properties. The success of SMM demonstrates that the heavy residues originate 
from events where a competition of thermally equilibrated fragment partitions 
takes place rather than a sequence of binary decays. 

\end{abstract}


\end{frontmatter}

\section{Introduction}

The yields of the heavy residues, the large
remnants of the heavy member of an asymmetric reacting pair of nuclei, are
known to comprise a large fraction of the reaction cross section for
intermediate energy nuclear collisions. 
The studies carried out in inverse kinematics at projectile energies 
of several tens of AMeV by Bazin et al.\cite{bazin}, Faure-Ramstein 
et al. \cite{faure}, Pfaff et al. \cite{pfaff}, Hanold et al. \cite{karl} 
and Souliotis et al. \cite{george,george2,george3} 
have shown the utility of the fragment separator 
approach in studying the heavy reaction products from Kr+X, Xe+X and Au+X mass 
asymmetric collisions at intermediate energies. 
Compared to asymmetric reactions in normal kinematics \cite{spectra,sarar}, 
where the low energies of residues ($\sim$0.015 AMeV) cause 
the loss of substantial portions ($\>$50\%) of the product distributions, 
due to experimental thresholds \cite{spectra},  
the studies in inverse kinematics using a high resolution 
spectrometer/detector system allow the observation of a wide range 
of nuclei including the projectile-like fragments (PLFs) 
and the heavy remnants of the hot nuclei originating from violent 
collisions. From such inclusive measurement, 
one can gain important information about the reaction mechanism, 
complementary to that obtained in exclusive studies,  where only the light
reaction partners are observed with high isotopic resolution.

For asymmetric reactions of a heavy beam with a light target nucleus 
( typically C, Al ) the momentum transfer data obtained 
( see e.g. \cite{faure,karl,george} ) demonstrate the presence of reaction 
mechanisms 
ranging from quasi-elastic peripheral collisions to processes analogous to the 
complete fusion and incomplete fusion observed at low energies. 
However, since asymmetric reactions at several tens of AMeV are 
complex processes, the description of the data by the simple low energy 
concepts such as massive-transfer hypothesis or the high-energy 
geometric abrasion model have shown to be, despite qualitative agreement, 
far from quantitative. 

With increasing target mass, the reactions become more symmetric and 
the observed yields of heavy residues are dominated by peripheral processes 
such as deep-inelastic transfer as, for instance, shown in detail 
in our recent work \cite{GSKrNi} in the reaction 
of 25 AMeV $^{86}$Kr + $^{64}$Ni. As it follows from the model analysis 
( for details see the original work ), the projectile-like fragments 
observed in this reaction are heated by intense exchange of nucleons with 
the target. The excitation energy is sufficient for emission of complex 
fragments. Previous studies 
\cite{bazin,granier,barz,yokoyama,marchetti,aboufirassi,garcia,baldwin,morjean,enterria} 
of the collisions of heavy nuclei at
intermediate energies have also shown that such reactions are 
predominantly binary. \ Evidence has been presented \cite{enterria}
for a sequential decay of one of the initial binary fragments leading to a
three (or more) body final state. \ In some of the binary encounters the
projectile-like fragments have been found \cite{aboufirassi,morjean} to have
very high temperatures (T$\sim $7 MeV). 

The focus of the present work is the formation of heavy residues 
in the reaction of 20 AMeV $^{124}$Sn on $^{27}$Al. The main motivation 
was the observation of heavy remnants of hot nuclei created 
in violent collisions at low impact parameters where a single hot 
source is created. In particular, it is of importance to establish 
in detail to what extent the observed properties 
of such heavy residues are influenced by the dynamics of the 
entrance channel ( pre-equilibrium emission, incomplete fusion ) 
and/or by the process of de-excitation ( emission of complex fragments, 
multifragmentation ). In the reaction $^{124}$Sn + $^{27}$Al one can 
assume that the residues with masses much lower than that of the beam originate 
from violent collisions. Such an assumption is supported by the work presented 
in \cite{MVSiSn}, where it was shown that for the damped peripheral collisions 
$^{28}$Si + $^{112,124}$Sn at energies 30 -- 50 AMeV the heavy target-like 
fragment remains rather cold, while the light projectile-like 
fragment is hot enough to undergo multifragmentation. This is caused 
by approximately equal sharing of the excitation energy imparted to the 
reaction partners due to 
nucleon exchange. In order to heat the heavy fragment to higher excitation
energies, part of the lighter partner should fuse with the heavier one, 
thus converting the kinetic energy of the relative motion into heat. 
The recoil spectrometer MARS at the Cyclotron Institute of Texas A\&M 
University offers the possibility to carry out this study with 
an appropriate angular and momentum acceptance, under high-resolution 
conditions and in the appropriate region of nuclei.

\section{Experimental Method and Data Analysis}

The present study was performed at the Cyclotron Institute of Texas A\&M
University. A  20 AMeV  $^{124}$Sn$^{27+}$ beam from the K500 superconducting 
cyclotron, with a typical current of $\sim$0.5 pnA, interacted with 
a $^{27}$Al target of  thickness 2.0 mg/cm$^{2}$. 
The reaction products were analyzed with  the MARS recoil separator \cite{MARS}.
The primary beam struck the target at 0$^{o}$ relative to the optical 
axis of the spectrometer.  The direct beam was collected in  a small square 
Faraday cup approx. 30 cm after the target,  blocking the angular range 0.0--1.0$^{o}$.
The fragments were accepted in the remaining angular opening of MARS:  1.0--2.7$^{o}$ 
(the angular acceptance of MARS is 9 msr \cite{MARS}). 
This angular range  encompasses the grazing angle of 2.1$^{o}$  \cite{Wilcke} 
for the present reaction. 
MARS optics \cite{MARS} provides one  intermediate dispersive  image and a 
final achromatic image (focal plane) and offers a  momentum acceptance of  
4\%.  

At the focal plane, 
the fragments were collected in a large area (5$\times$5 cm) three-element 
($\Delta $E$_{1}$,  $\Delta $E$_{2}$, E) Si detector telescope. 
The $\Delta$E$_{1}$ detector was a position-sensitive Si strip detector 
of 63 $\mu$m 
thickness,  whereas the $\Delta$E$_{2}$ and the E detector were  
single-element Si detectors of
150 and  950 $\mu$m, respectively.
The position information from the $\Delta $E$_{1}$ strips provided a continuous 
monitoring of the focusing and collection of the fragments at the various 
settings of the separator.
Time of flight was measured between two parallel plate avalanche 
counters (PPACs) \cite{Greg}
positioned at the dispersive image and at the focal plane, respectively, 
and separated by a distance of 13.2  m. 
The PPAC at the dispersive image was also  X--Y  position sensitive  and  
used  to record 
the position of the reaction products. The horizontal position, along with NMR
measurements of the field of the MARS first dipole, 
was used to determine the magnetic rigidity $B\rho $ of the particles. 
Thus, the reaction products were characterized by an event-by-event measurement
of the energy loss, residual energy, time of flight, and magnetic rigidity. 
The response of the spectrometer and detector system 
to ions of known atomic number Z, mass number A, ionic charge q and 
velocity was calibrated using low intensity primary beams of 
$^{124}$Sn at 20 AMeV and $^{40}$Ar, $^{44}$Ca  and  $^{86}$Kr at 25 AMeV. 
To cover the N/Z and velocity range of the fragments, a series of measurements
was performed at overlapping magnetic rigidity settings in the range 1.3--1.6 
Tesla-meters. 

The determination of the atomic number Z was based on the energy loss of the 
particles in the first $\Delta E$ detector \cite{Hubert} and their velocity,
with a resulting resolution (FWHM) of 1.0 Z units for near-projectile
fragments and 0.6 Z units for A$<$90. 
The ionic charge $q$ of the particles entering the spectrometer 
after the Al stripper, was obtained from
the total energy E$_{tot}$=$\Delta$E$_1$+$\Delta$E$_2$+E, the velocity and 
the magnetic rigidity
according to the expression: 
\begin{equation}
q=\frac{3.107}{931.5}\frac{E_{tot}}{B\rho (\gamma -1)}\beta \gamma
\label{q_eqn}
\end{equation}
where E$_{tot}$ is in MeV, B$\rho $ in Tm, $\beta =\upsilon /c$ and $\gamma
=1/(1-\beta ^2)^{\frac 12}$. 
The measurement of the ionic charge q had a resolution of 0.8 Q units (FWHM) 
for near-projectile fragments and 0.5 Q units for A$<$90. 
Since the ionic charge must be an integer, we assigned integer
values of q for each event by putting appropriate windows  
on each peak of the q spectrum at each magnetic rigidity setting of the 
spectrometer.
Using the magnetic rigidity and velocity measurement, the mass-to-charge 
A/q ratio  of each ion was obtained from the expression: 
\begin{equation}
A/q = \frac{B\rho }{3.107\beta \gamma }  \label{Aq_eqn}
\end{equation}
Combining the q determination with the A/q measurement, the mass A
was obtained as:
\begin{equation}
A = q_{int} \times A/q  \label{A_eqn}
\end{equation}
(q$_{int}$ is the integer ionic charge determined as above) with an 
overall resolution  (FWHM) of 1.0 A units for near-projectile fragments and 
about 0.6 A units for A $<$ 90. We refer to our previous work in 
ref. \cite{GSKrNi}, carried out using the same 
experimental setup within the same run, for more details. 
The reconstruction of the Z, q and A and the gating procedure 
were applied to the calibration
beam data to ensure the reproduction of the expected Z, q and A values and the 
elimination of spurious yield contributions from neighboring Z and q values.
For the heavier products with masses A $>$ 90 the experimental resolutions 
did not allow the mass to be resolved unambiguously. However, the gross 
features for such products can still be obtained. Close to the beam, part 
of the yield could not be detected due to background from the quasi-elastically 
scattered beam.


Combination and appropriate normalization of the data at various magnetic
rigidity settings of the spectrometer provided fragment  distributions with 
respect to 
Z, A, q and velocity. Correction of missing yields caused by charge changing 
at the  PPAC (positioned at the dispersive image) was performed
based on the equilibrium charge state prescriptions of Leon et. al. \cite{Leon}. 
The overall data reduction procedure was similar to that followed in earlier 
work on $^{197}$Au fragmentation \cite{george2}
and $^{238}$U projectile fission \cite{GSfission} at 20 AMeV.
The isotope distributions were subsequently summed over all values of q. 
It should be pointed out  that the resulting distributions in Z, A and velocity 
are the fragment yield distributions measured 
in the angle interval 1.0--2.7$^{o}$ and 
in the magnetic rigidity range 1.3--1.6 T\,m.

\section{Results and Discussion}

\begin{figure}[p]                    

\includegraphics[width=10.cm,height=10.cm]{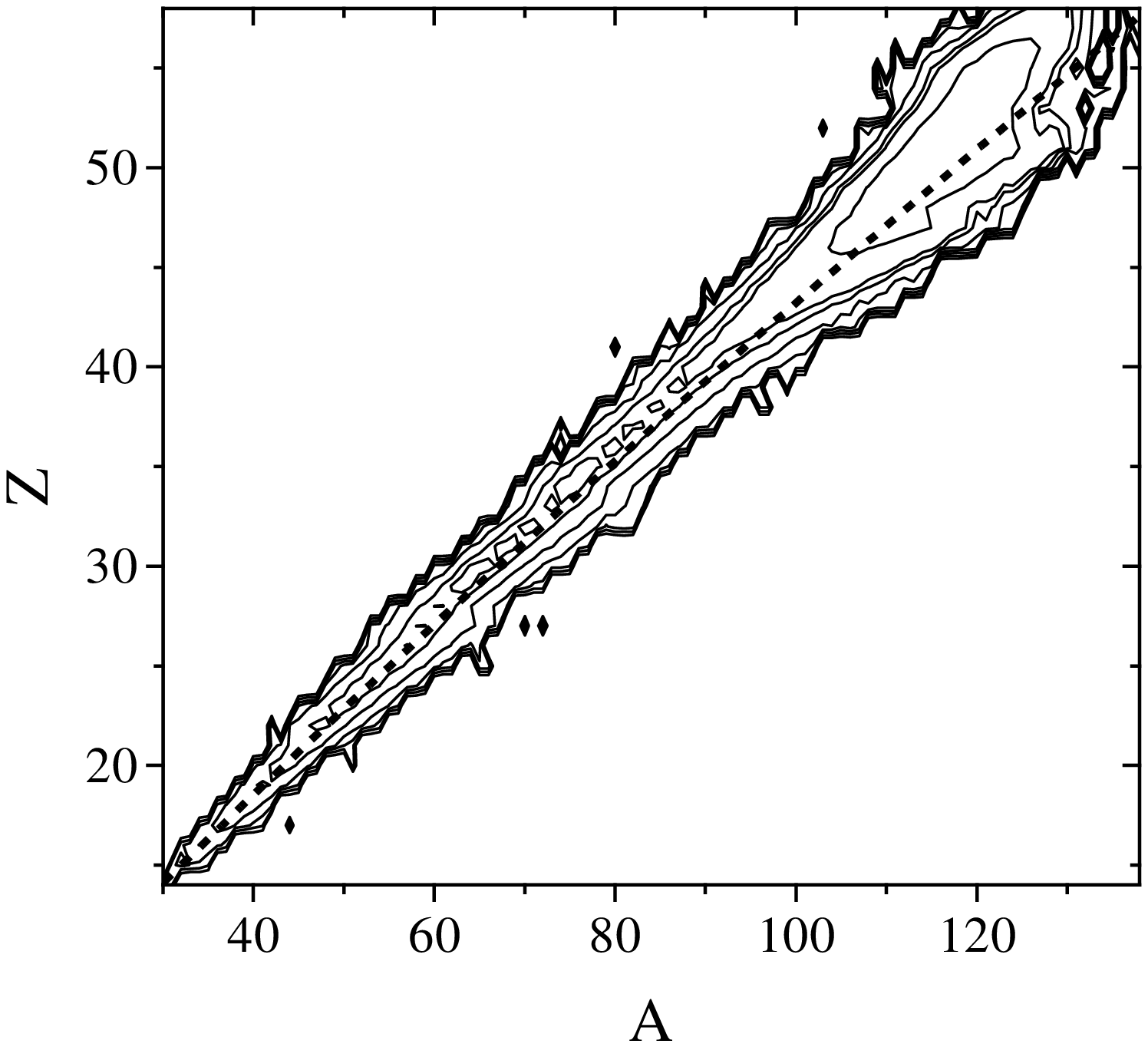}
\centering
\vspace{0.0cm}

\caption{ 
Isotopic yields ( contour plot ) for the reaction  
20 AMeV  $^{124}$Sn + $^{27}$Al as a function of mass and atomic number. 
Successive contours correspond to a decrease  of the yield by a factor of 10.
The corridor of stable isotopes is given by a thick dashed line.
   }

\label{yaz}

\end{figure}

\begin{figure}[p]                    

\includegraphics[width=10.cm,height=10.cm]{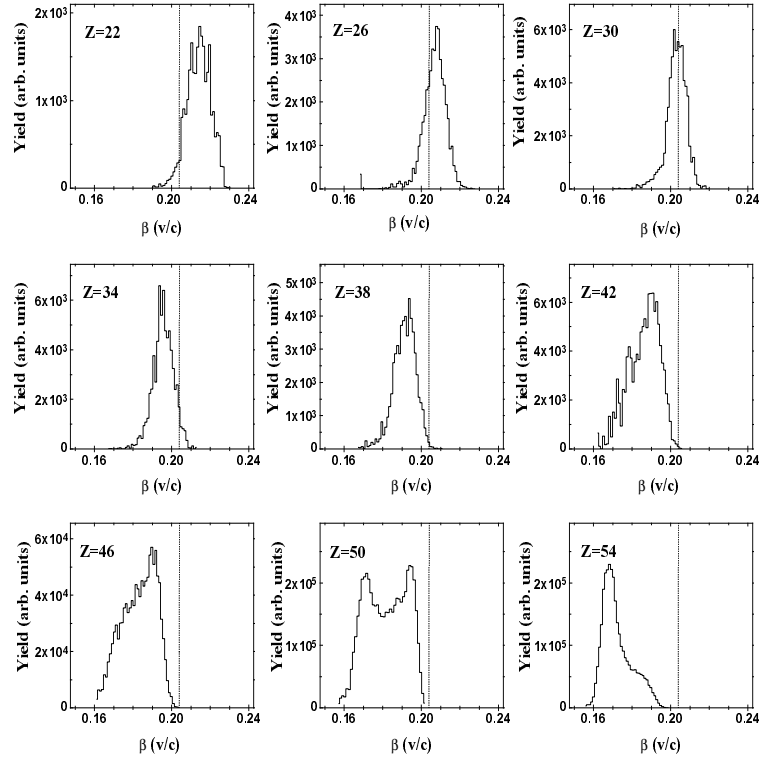}
\centering
\vspace{0.0cm}

\caption{ 
Velocity distributions of residues with selected atomic numbers measured 
in the reaction  20 AMeV  $^{124}$Sn + $^{27}$Al in the laboratory system. 
The vertical dotted line represents the beam velocity.    }

\label{vdist}

\end{figure}

The gross features of the measured data from the reaction 
of 20 AMeV $^{124}$Sn + $^{27}$Al are displayed in Figs. \ref{yaz},\ref{vdist}. 
In Fig. \ref{yaz} isotopic yields ( contour plot ) are presented as 
a function of mass and atomic number. 
The masses of observed products cover the range from 
the projectile-like fragments to the border of the intermediate mass 
fragment domain at Z=20. The yields of projectile-like nuclei peak 
around the nuclide $^{116}$Te. The position of this peak is caused by missing 
yield from the region close to the beam where the reaction products 
cannot be resolved using the existing detector set-up due to significant 
background from quasi-elastic processes. At masses below A=90 the distribution 
becomes rather flat and follows approximately the corridor of stable 
isotopes ( thick dashed line ). These are the reaction products which 
are expected to originate from the hot excited systems. 
Fig. \ref{vdist} presents the velocity distributions of residues with selected 
atomic numbers. In the region close to beam the velocity distributions 
appear to be convolutions of multiple contributions from different reaction 
channels ranging from quasielastic products close to the beam velocity to 
various incomplete fusion channels where the velocity decreases with the amount 
of mass transferred from the target. With decreasing mass of the reaction 
product, the velocity distributions develop toward a single contribution 
of a Gaussian shape with a low velocity tail. The centroids 
of these distributions shift with decreasing mass toward higher velocity. 
Such a trend is caused by the kinematic selection of the spectrometer. 

The experimental heavy residue data presented in Figs. \ref{yaz},\ref{vdist} 
contain useful information on the production mechanism. 
The properties of the final products are the result of a complex process 
including an early stage dominated by the dynamics of the entrance channel 
which is governed by the impact parameter and a de-excitation stage where the 
final partitions of the reaction products are generated. 
In such a case, a viable method of reaction mechanism analysis 
appears to be the use of various model frameworks for both the initial stage 
of the collision and the de-excitation. The model which eventually 
proves superior to the others can be considered as reflecting the physical 
process in most detail. 

As a first choice for the description of the initial stage we 
use the model framework described in \cite{martin2}.  
The basic features of the reaction mechanism model for violent collisions 
are the pre-equilibrium emission and the incomplete fusion (ICF). 
The pre-equilibrium emission is treated using a simplified variant of the 
exciton model employing a phenomenological parametrization of emission 
probability as a function of exciton number and angular momentum. 
The incomplete fusion model is based on the concept of geometrical 
fragmentation refined for the Fermi energy domain where incomplete fusion 
occurs by fusion of the participant zone with one of the spectators. 
It is applied to the reconstructed projectile-like and target-like prefragments 
formed in the pre-equilibrium stage. For the dissipative peripheral collisions 
the model of deep inelastic transfer (DIT) is used as implemented 
by Tassan-Got and Stefan \cite{tassan}. The DIT scenario is employed 
for events where the overlap of projectile and target nuclei does not 
exceed 3 fm. Again the DIT stage is preceded by the pre-equilibrium stage. 
Such a hybrid framework proved rather successful \cite{martin2} in 
the description of a wide range of data obtained in experiments ranging 
from inclusive measurements to highly exclusive ones in 4$\pi$ geometry 
and can be considered appropriate for the reaction investigated 
in the present work, with a capability of quantitative description. 
Throughout the paper, this calculation will be named 
for simplicity as DIT+ICF. Nevertheless, one has to keep in mind that 
pre-equilibrium emission is an essential part of this model. 

In order to describe the de-excitation stage of hot nuclei, there exist 
several concepts implemented in various codes. The statistical de-excitation 
code GEMINI \cite{GEMINI} uses Monte Carlo techniques 
and the Hauser-Feshbach formalism to calculate the probabilities for 
fragment emission with Z$\leq$2. Heavier fragment emission 
probabilities are  calculated using the transition state formalism of 
Moretto \cite{Moretto}. Within such a model, the final partition 
of products is generated by a succession of fragment emissions 
( binary decays ). Alternative to this scenario is the model 
of statistical multifragmentation, where the fragment partition 
is generated at once in the so-called freeze-out configuration. 
In this work, we use the code SMM \cite{smm} as a representative 
implementation of the concept of prompt multifragmentation. 
Thermally equilibrated partitions of hot fragments are generated 
in the hot stage, which is followed by propagation of fragments in their 
mutual Coulomb field and secondary de-excitation of hot fragments 
flying in their asymptotic directions. 

\begin{figure}[p]                    

\includegraphics[width=10.1cm,height=15.5cm]{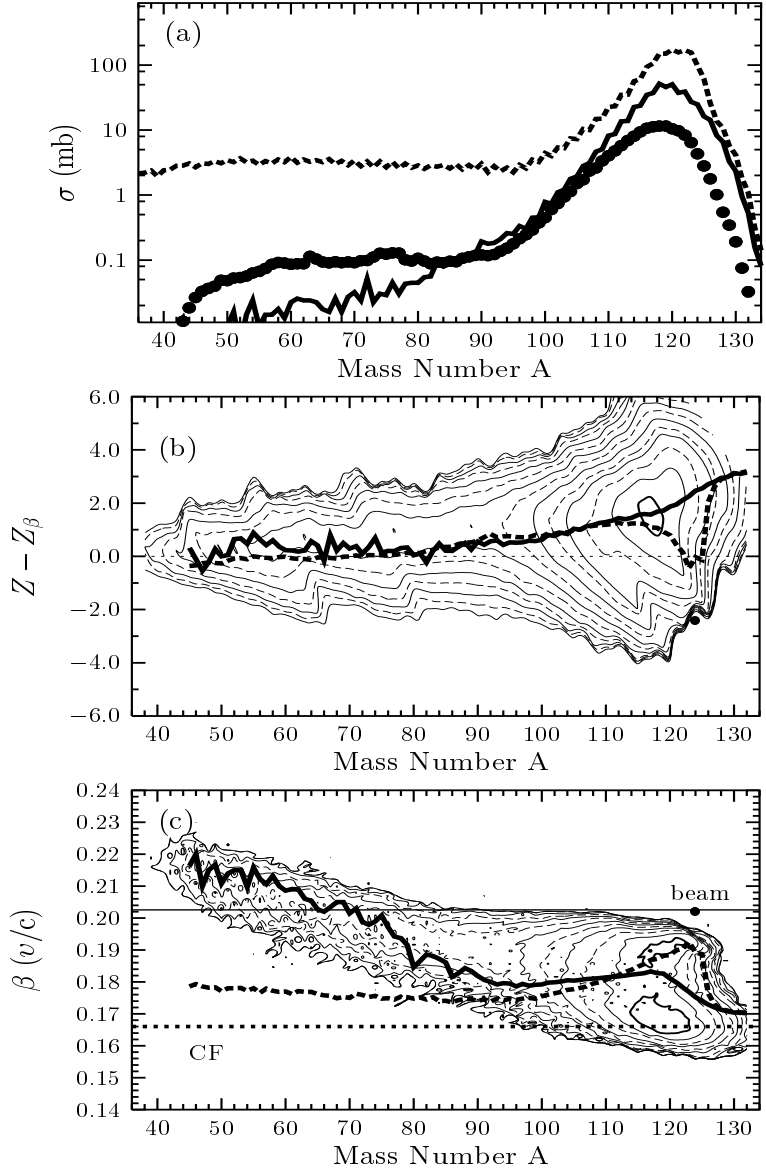}
\centering
\vspace{0.0cm}

\caption{ 
    Fragment distributions for the reaction  
    20 AMeV  $^{124}$Sn + $^{27}$Al. 
(a) - isobaric yield distribution. The data are shown as solid circles. 
    The dashed line is the result of DIT+ICF/GEMINI (see text). 
    The full line is  the result of the  same calculation as the dashed line, 
    but imposing a cut corresponding  to the angular and momentum acceptance 
    of the spectrometer.  
(b) - yield distributions 
    as a function of Z (relative to the line of $\protect\beta$ stability,
    Z$_{\protect\beta}$) and A.
    Highest yield contours are plotted with thicker lines. Successive contours
    correspond to a decrease  of the yield by a factor of 2.
    The calculated values from DIT+ICF/GEMINI are shown as i) thick dashed line:
    without acceptance cut and, ii) thick full line: with acceptance cut.
(c) - velocity vs. mass distributions.
    Data are shown as contours as in (b). The thick lines are as in (b).
    The  horizontal full line represents the  beam velocity and the horizontal 
    dashed line represents the velocity of compound nucleus in the case of 
    complete fusion.
   }

\label{ygem}

\end{figure}

The results of the DIT+ICF/GEMINI calculation, compared to experimental 
observables are given in Fig. \ref{ygem}.  
In the GEMINI calculations, we used essentially the parameter set 
recommended by the author, featuring Lestone's temperature dependent 
level density parameter \cite{Lestone}, a fading of shell corrections with
excitation energy and enabled IMF emission. This parameter set 
proved successful in our recent work \cite{GSKrNi} on the reaction 
$^{86}$Kr+$^{64}$Ni at 25 AMeV.

In Fig. \ref{ygem}a the mass yield curve is presented.
The measured data, normalized for beam  current and target thickness 
are given in mb and presented as solid circles.
The result of the DIT+ICF/GEMINI calculation, filtered by the spectrometer
angular and momentum acceptance is given by the full line, whereas the dashed 
line gives the total (unfiltered) yield. 
A comparison of the measured yields to the calculated filtered yields
shows reasonable agreement for the heavier projectile-like 
fragments (A$>$90). The missing experimental cross section 
at the masses close to the beam is caused by the limitations 
of the experimental set-up as explained in the discussion of Fig. \ref{yaz}. 
The yields of residues with A$<$90 are increasingly underestimated 
by the calculation despite the fact that the unfiltered calculated yields 
are rather flat in this region, which appears to reflect the trend 
of the experimental data. 
Since, according to Fig. \ref{ygem}a, the yield of unfiltered residues 
with A$<$90 is rather flat, 
the missing filtered yield below A=90 in the calculation appears to be 
caused by the fact that the kinematic properties of the simulated 
residues with A$<$90 increasingly miss the spectrometer acceptance 
with decreasing mass, leading to increasing losses in the filtering procedure. 

In Fig. \ref{ygem}b,  the calculated and measured yield distributions 
as a function of Z (relative to the line of $\protect\beta$-stability,
Z$_{\protect\beta}$) and A are presented. 
The line of stability is calculated as: 
Z$_{\beta }$ = A/(1.98 + 0.0155A$^{2/3}$) \cite{Marmier}.
The calculated average values from DIT+ICF/GEMINI are shown as  a thick 
dashed line 
(without acceptance cut) and  as a thick full line (with spectrometer 
acceptance cut). The measured yield distribution is represented 
by a contour plot. 
The DIT+ICF/GEMINI calculation describes reasonably well the centroids 
of the experimental data.

Finally, in Fig. \ref{ygem}c, the velocity vs. mass distributions 
are given. The data  are again shown as contours. 
The thick dashed line is from the DIT+ICF/GEMINI calculation without 
acceptance cut and the full line is with  acceptance cut. 
In this case the filtered calculated data appear to follow 
the experimental trend at masses above A=90, while they exhibit 
slight increase of the mean velocity with decreasing 
masses below A=90. 

\begin{figure}[p]                    

\includegraphics[width=10.1cm,height=15.5cm]{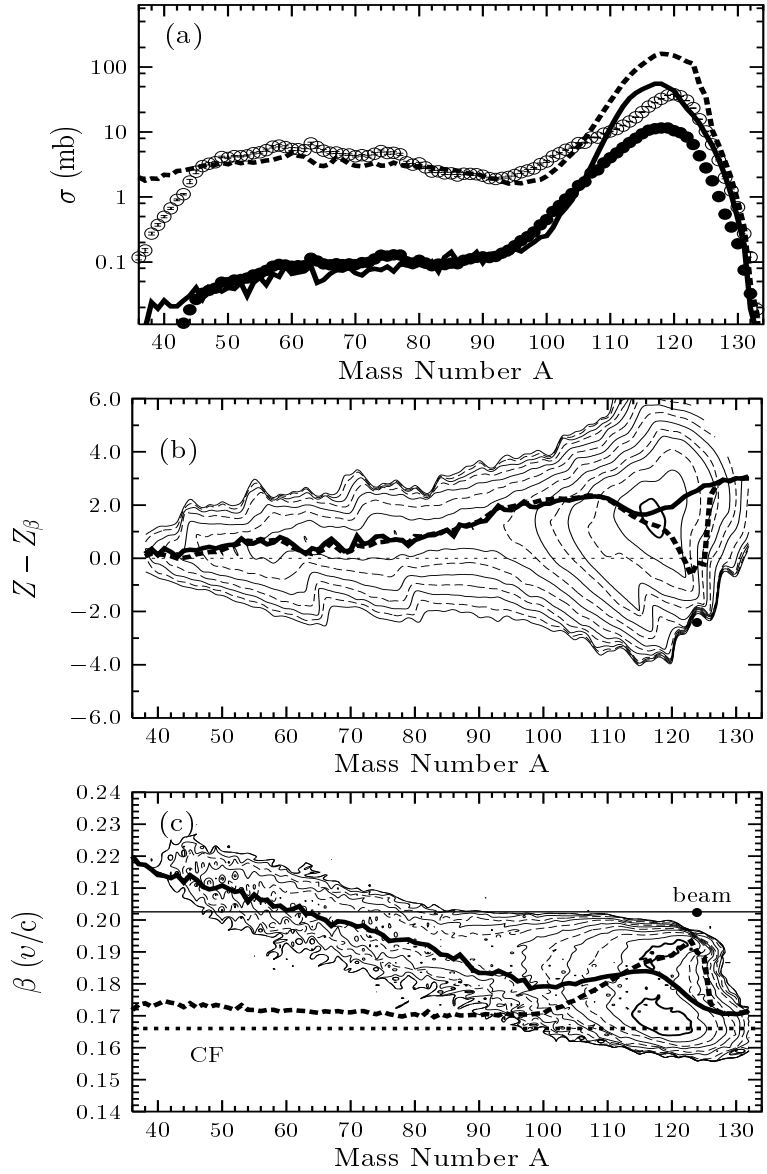}
\centering
\vspace{0.0cm}

\caption{ 
    Fragment distributions for the reaction  
    20 AMeV  $^{124}$Sn + $^{27}$Al as in Fig. \ref{ygem}, except that 
    the calculations are DIT+ICF/SMM. Open circles in (a) 
    show estimated total cross sections (see text). 
   }

\label{ysmm}

\end{figure}

An alternative calculation was carried out, where the SMM code was used 
for the de-excitation of the hot nuclei produced by the DIT+ICF simulation. 
In the SMM calculation, a freeze-out configuration with hot primary fragments 
was assumed. Hot fragments are propagated in the Coulomb field and 
de-excited by secondary emission. Only thermal excitation energy is used 
as input while the rotational energy ( typically not exceeding 10 MeV ) is 
subtracted. The results are presented in Fig. \ref{ysmm} in a fashion analogous 
to Fig. \ref{ygem}. As one can see in all panels of Fig.\ref{ysmm}, 
the DIT+ICF/SMM calculation provides very consistent description 
of experimental observables for A$<$105 ( the discrepancies in the region 
close to the beam are analogous to previous cases ). 
Using the ratio of filtered to unfiltered calculated yield for each mass,
correction factors (whose magnitude are inferred from Fig. \ref{ysmm}a) 
for the acceptance of the spectrometer
can be obtained as a function of mass.  When applied to the measured yield 
data, an estimate of the total yield, shown by the open circles 
in Fig. \ref{ysmm}a, could be obtained. 

Thus, implementation 
of the prompt multifragmentation scenario appears to lead to production 
of heavy residues with proper kinematics. 
Compared to light particles or intermediate mass fragments (IMFs) 
used for imaging of the emitting 
source via particle-particle correlations, the experimentally detected 
heavy residues possess direct information on the properties of hot 
multifragment partitions. The process of secondary 
emission, as can be concluded from the simulations, does not influence 
significantly the kinematic properties of the heavy residues with 
masses A=40--90, since emission of nucleons is a dominant channel 
of secondary de-excitation. 

\begin{figure}[p]                    

\includegraphics[width=10.1cm,height=15.5cm]{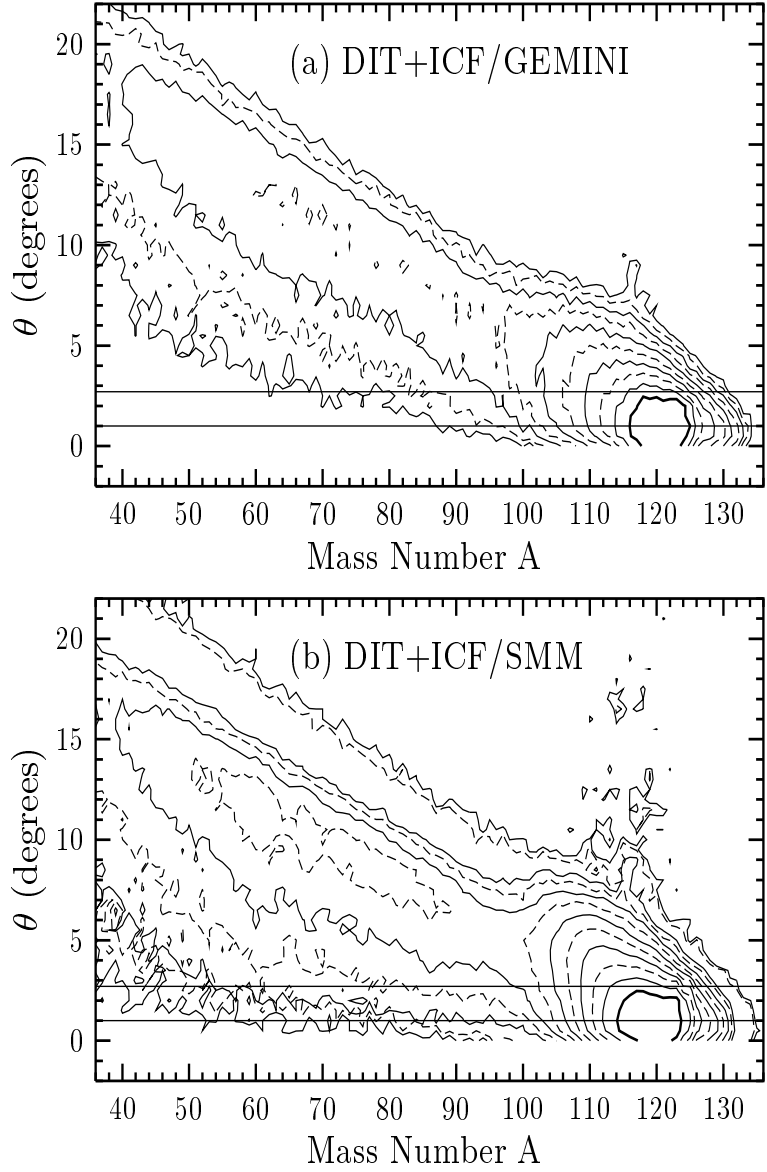}
\centering
\vspace{0.0cm}

\caption{ 
Calculated angular distributions in the laboratory system 
for the reaction  20 AMeV  $^{124}$Sn + $^{27}$Al 
as a function of residue mass for GEMINI (a) and SMM (b). 
Successive contours correspond to a decrease  of the yield by a factor of 2.
The two horizontal lines mark the angular acceptance of the MARS separator 
used in the present work.    
}

\label{thgemsmm}

\end{figure}

\begin{figure}[p]                    

\includegraphics[width=10.1cm,height=15.5cm]{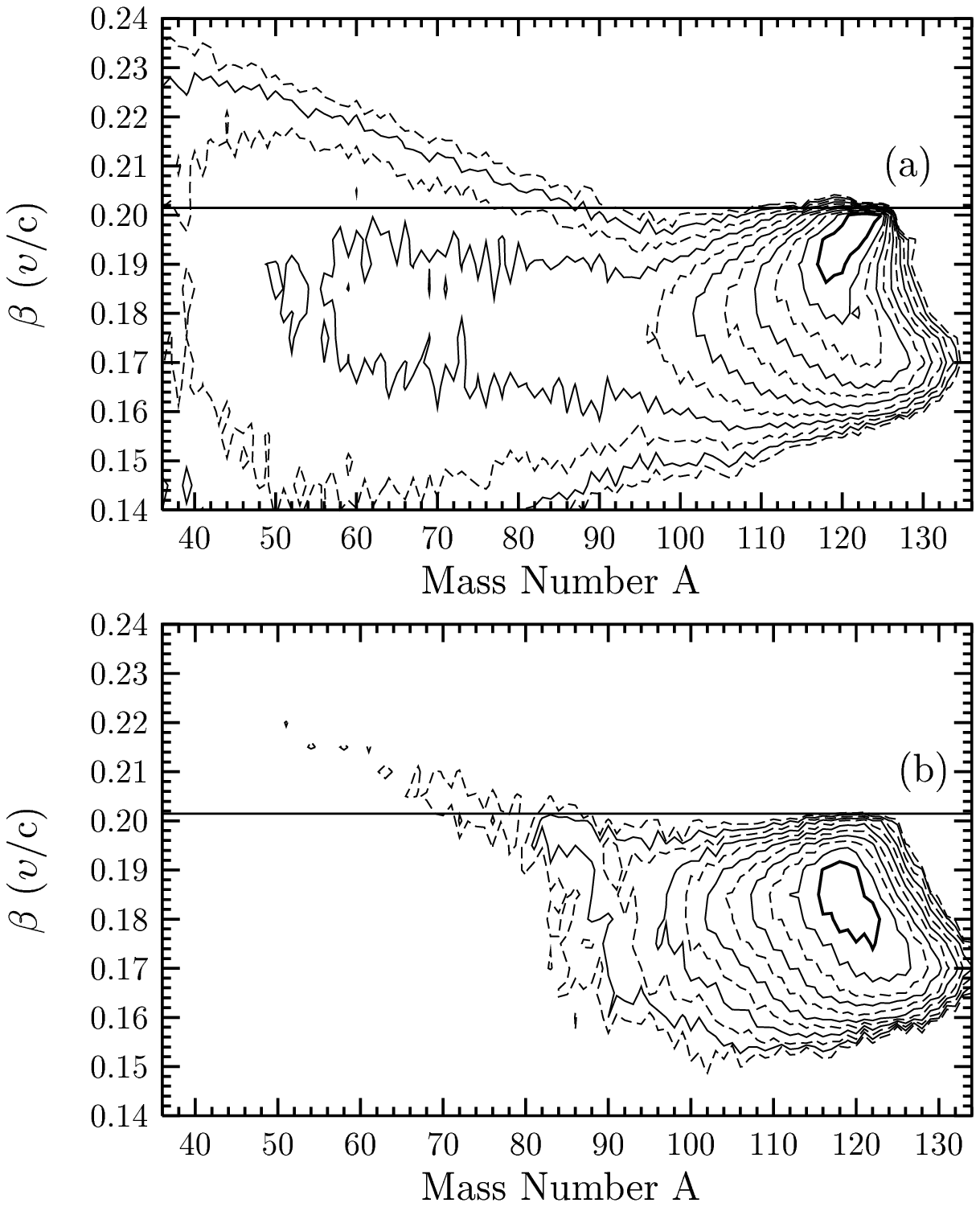}
\centering
\vspace{0.0cm}

\caption{ 
Calculated velocity distributions 
for the reaction  20 AMeV  $^{124}$Sn + $^{27}$Al 
as a function of mass obtained using the DIT+ICF/GEMINI simulation. 
Unfiltered yields are presented in panel (a), while the filtered 
yields are given in panel (b). Successive contours correspond to 
a decrease  of the yield by a factor of 2.
The horizontal lines mark the beam velocity.    
}

\label{vgem}

\end{figure}

\begin{figure}[p]                    

\includegraphics[width=10.1cm,height=15.5cm]{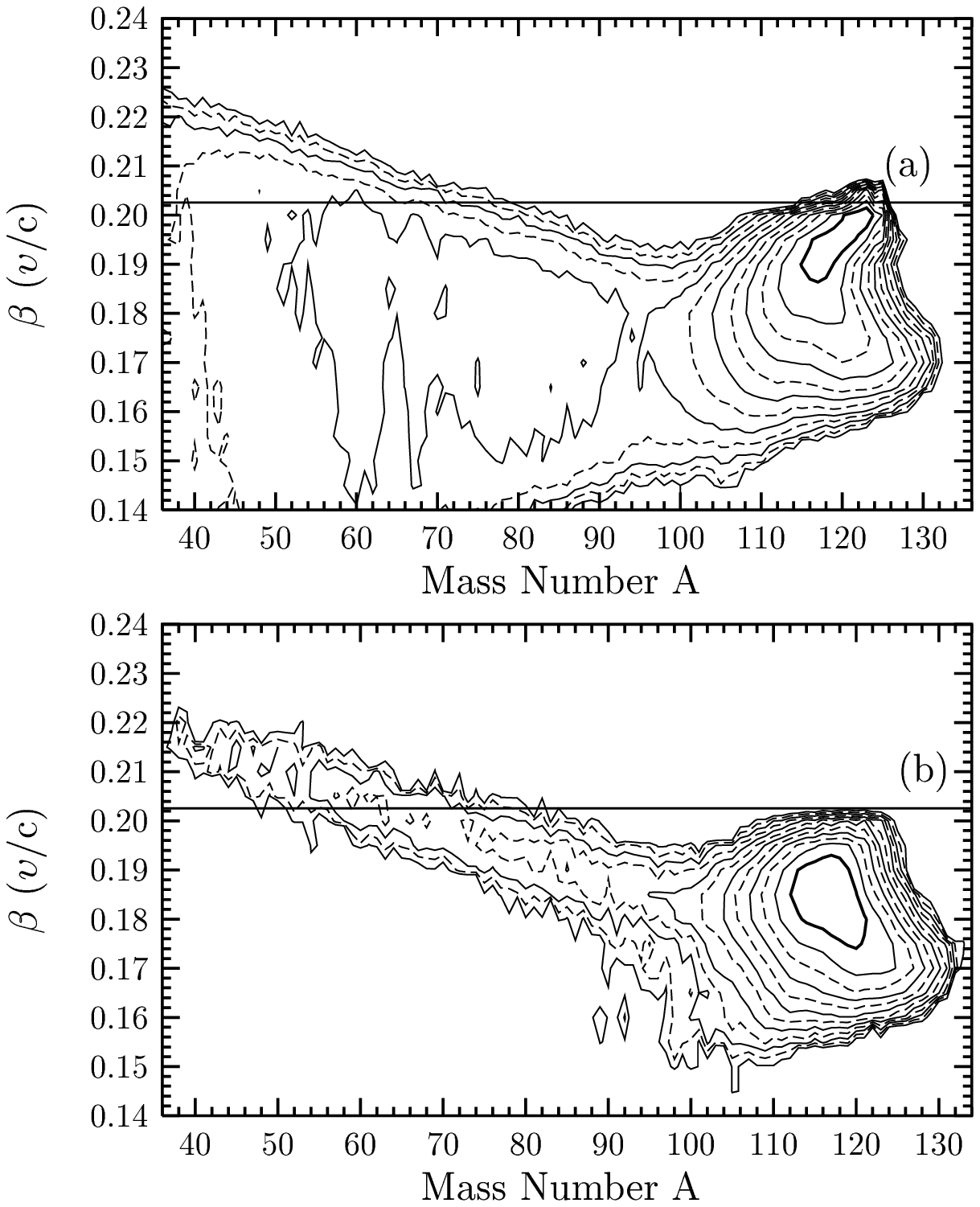}
\centering
\vspace{0.0cm}

\caption{ 
Calculated velocity distributions 
for the reaction  20 AMeV  $^{124}$Sn + $^{27}$Al 
as a function of mass obtained using the DIT+ICF/SMM simulation. 
Unfiltered yields are presented in panel (a), while the filtered 
yields are given in panel (b). Successive contours correspond to 
a decrease  of the yield by a factor of 2.
The horizontal lines mark the beam velocity.    
}

\label{vsmm}

\end{figure}

Detailed insight into the different kinematic properties of residues simulated 
using GEMINI and SMM codes can be obtained from 
Figs. \ref{thgemsmm},\ref{vgem} and \ref{vsmm}. 
In Fig. \ref{thgemsmm} the calculated angular distributions as a function 
of residue mass for GEMINI (a) and SMM (b) are presented. 
Two horizontal lines mark the angular acceptance of the MARS separator. 
It is remarkable to notice that, in fact, the gross features of both 
distributions are very similar and the experimental effect appears 
to be caused by the distant tail of the distribution, which extends 
much further toward zero angle in the scenario where the hot nucleus 
disintegrates at once into more pieces. Nevertheless, the essential feature of 
the zero angle region is that it is highly selective toward products 
from incomplete or complete fusion channel since the hot source is flying 
essentially along the beam direction and the final angular distribution is 
determined by the de-excitation process. In the case when the hot nucleus 
disintegrates by sequential binary decay, the recoil from emitted fragments 
causes a shift of the residue angle away from the zero angle. 
In Figs. \ref{vgem},\ref{vsmm} the calculated velocity distributions 
as a function of mass obtained using the GEMINI and SMM codes are given, 
respectively. 
In both cases, unfiltered yields are presented in panel (a), while the filtered 
yields are given in panel (b). As one can observe in panel (a), the 
SMM calculation leads to a much wider velocity distribution for A$<$90 
which corresponds to a flatter angular distribution extending further toward 
zero angle. In the GEMINI calculation, the velocity distribution is 
concentrated close 
to the center of the distribution for a given mass, while the tails toward 
higher/lower velocities are suppressed since such residues, due to 
kinematics, can be produced only close to zero angle. On the other hand, 
when comparing the fastest residues for a given mass at A$<$90, 
which are emitted close to zero angle, the GEMINI calculation leads to higher 
maximum velocities, due to a larger recoil caused by the scenario of sequential 
binary decay. Nevertheless, due to the momentum coverage of the whole 
measurement which is close 
to 20 \%, such a difference does not have a strong influence on the 
filtered yield. The effect of angular acceptance of MARS is dominant 
and determines the filtered yields which are much higher in the SMM simulation 
than in the GEMINI simulation.

\begin{figure}[p]                    

\includegraphics[width=10.1cm,height=15.5cm]{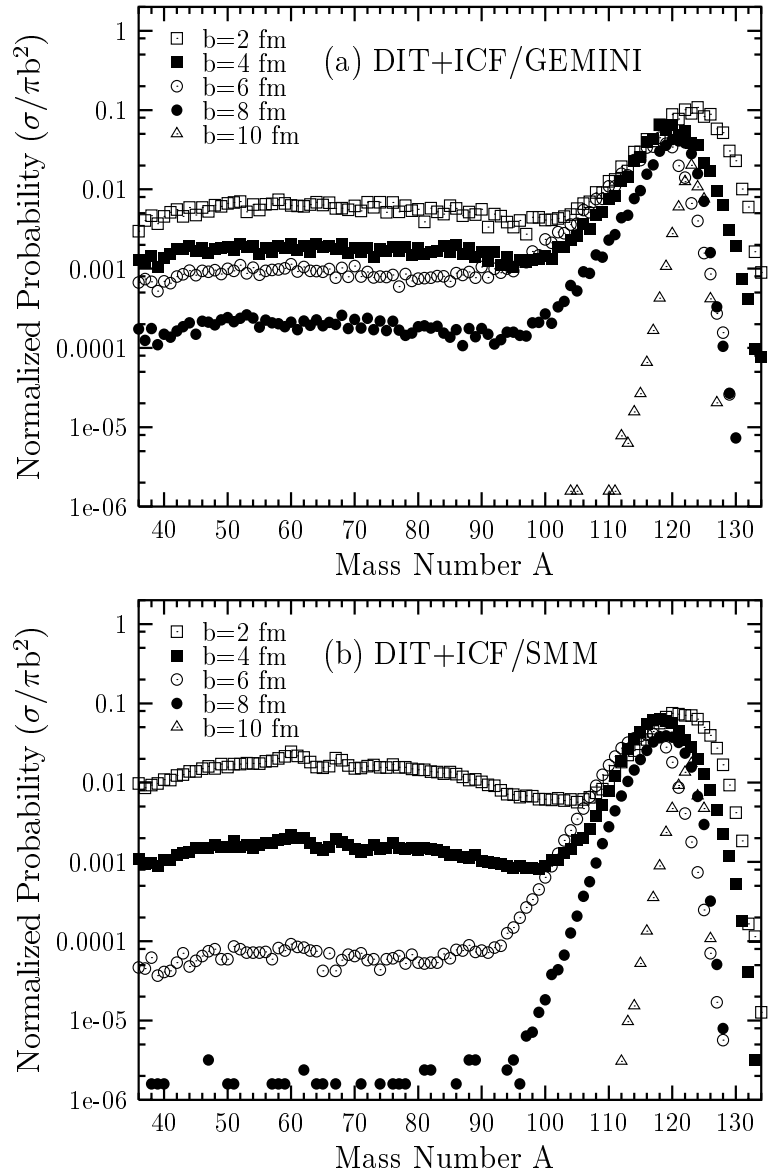}
\centering
\vspace{0.0cm}

\caption{ 
Normalized unfiltered residue mass distributions 
for the reaction  20 AMeV  $^{124}$Sn + $^{27}$Al 
at various impact parameters obtained using (a) DIT+ICF/GEMINI, 
and (b) DIT+ICF/SMM simulations. 
   }

\label{prob}

\end{figure}

\begin{figure}[p]                    

\includegraphics[width=10.1cm,height=15.5cm]{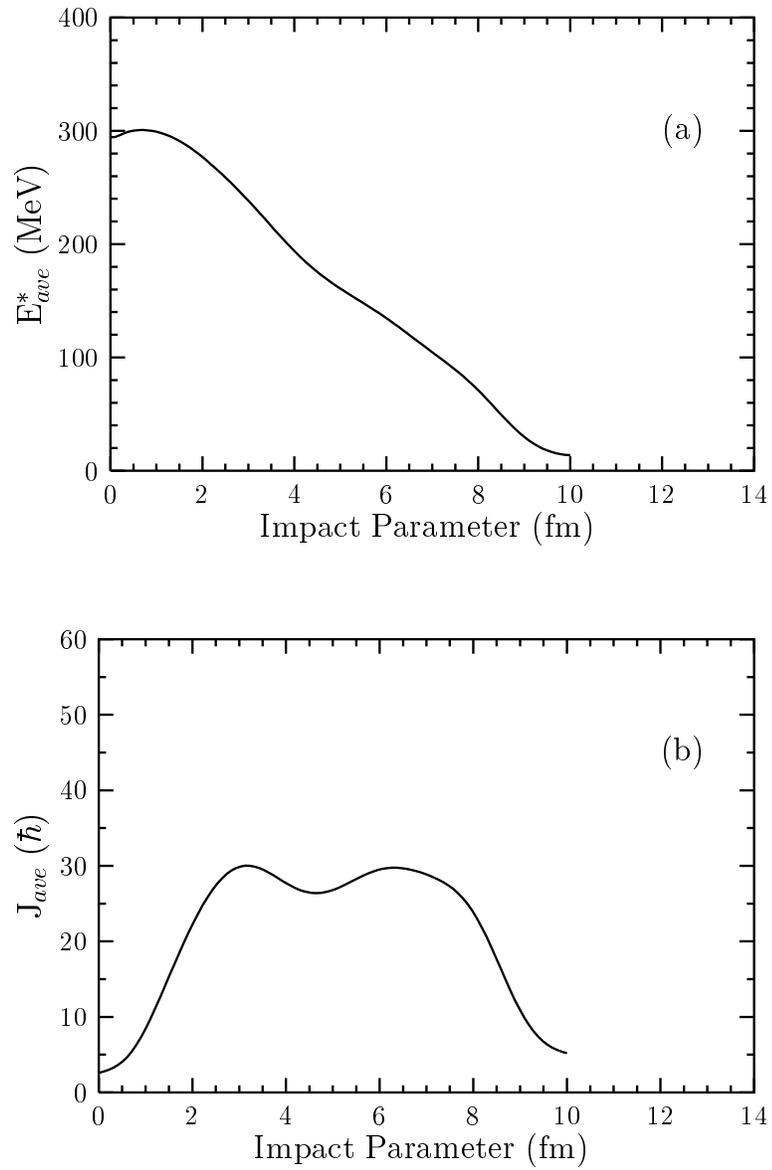}
\centering
\vspace{0.0cm}

\caption{ 
Average excitation energy (a) and angular momentum (b) of the heavy source 
as a function of impact parameter obtained for the reaction  
20 AMeV  $^{124}$Sn + $^{27}$Al using the DIT+ICF model. 
   }

\label{bdep}

\end{figure}

\begin{figure}[p]                    

\includegraphics[width=10.cm,height=10.cm]{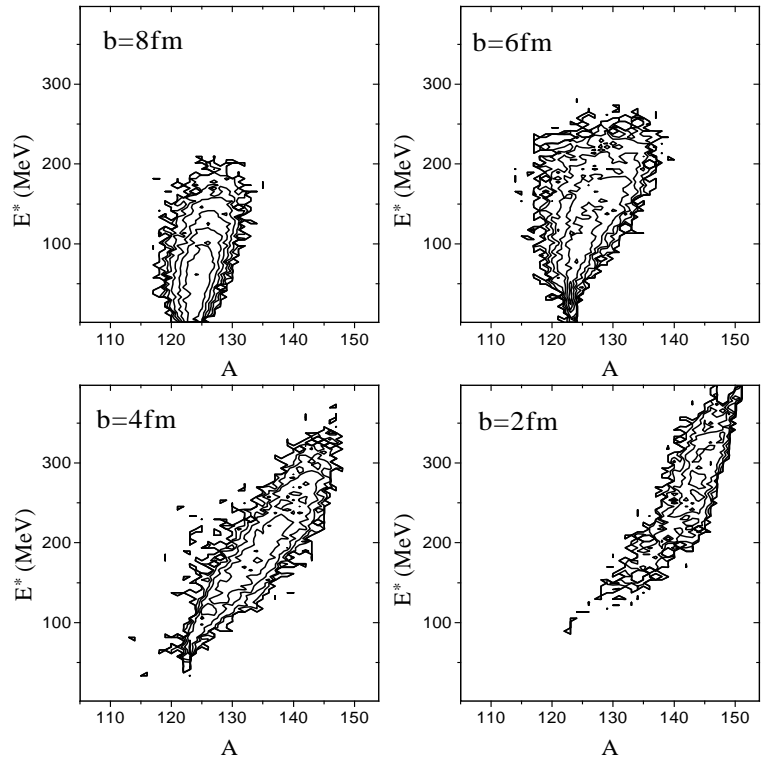}
\centering
\vspace{0.0cm}

\caption{ 
Calculated excitation energy vs mass distribution of the heavy source 
at various impact parameters obtained 
for the reaction  20 AMeV  $^{124}$Sn + $^{27}$Al 
using the DIT+ICF simulation. 
Successive contours correspond to a decrease  of the yield by a factor of 2.
   }

\label{excimp}

\end{figure}

In Fig. \ref{prob} the normalized unfiltered residue mass 
distributions at various impact parameters 
are given for both calculations. For DIT+ICF/SMM, one 
can notice a rather strong dependence of the production rate for residues 
with A$<$90 on impact parameter. For the DIT+ICF/GEMINI calculation, the 
dependence is rather weak. Significant contribution of A$<$90 residues 
is produced already at b=8 fm and toward lower impact parameters the 
shape of the distribution changes only moderately. As one can see 
in Fig. \ref{bdep}a, the average excitation energy of the hot source 
obtained using DIT+ICF model 
changes with impact parameter rather strongly. At b=8 fm, the average 
excitation energy amounts only to 80 MeV, which translates into relative 
excitation below 1 AMeV. At such excitation energy, the emission of 
IMFs is rather improbable and the simulated residues with A$<$90 can  
be expected to be fission fragments in the traditional sense. However, 
the mass distribution of such residues is virtually flat due to a 
weak asymmetry dependence of fission barriers of nuclei with fissility 
$x$ = 0.4 - 0.5, as suggested by theory \cite{SierkAsym}. 
In the GEMINI calculation, such fission fragments are produced with a 
probability of about two orders of magnitude larger than in the SMM 
calculation. This rather large difference in yield reflects a different 
treatment of fission in each code. The code GEMINI uses the 
asymmetry-dependent fission barriers \cite{SierkAsym} normalized 
to angular momentum dependent finite range fission barriers 
of Sierk \cite{sierk}. On the other hand, the SMM code, 
where fission and multifragmentation are treated separately, uses 
the fission barriers of Barashenkov \cite{barash} without explicit 
angular momentum dependence ( the mass distributions in the SMM are 
generated according to the parametrization of Adeev \cite{Adeev} ). 
For example, the nucleus $^{124}$Sn 
at angular momentum $J$=0 is supposed to have the fission barrier 
of 41.5 MeV in GEMINI and 44.8 MeV in SMM. At $J$=25, which is 
a typical value of angular momentum 
at b=8 fm ( see Fig. \ref{bdep}b ), the fission barrier of Sierk drops 
further to 36.9 MeV. Assuming a level density parameter $a$=A/10, 
the difference of level density indeed amounts to slightly less 
than two orders of magnitude, 
consistent with Fig. \ref{prob}. Thus the production rates 
of residues with A$<$90 at b=8 fm are consistent with the production rates of 
fission fragments. An analogous conclusion applies to impact parameter b=6 fm. 
At central impact parameters b=4 fm and b=2 fm, where the average relative 
excitation energy approaches and exceeds 2 AMeV, the 
production rate of residues with A$<$90 increases dramatically 
in the SMM calculation and exceeds the production rate of GEMINI calculation, 
since the multifragmentation threshold is exceeded by an increasing part 
of the excitation energy distributions at central impact parameters 
b=4 fm and b=2 fm, as one can deduce from Fig. \ref{excimp}.  

For the heavier asymmetric system $^{136}$Xe+$^{48}$Ti at 18.5 AMeV, 
where, due to higher fissility, the fission mass distribution is dominated 
by symmetric mass splits, the experimental investigation of 3-body events 
carried out by Gui et al. \cite{Gui} revealed that the sequential fission mode 
with symmetric mass distribution dominates at lower excitation energies, 
while at higher excitation energies the de-excitation mode with wider asymmetry 
range and short emission time was observed. The asymmetric de-excitation mode 
with short timescale can, in principle, be related to the onset of 
multifragmentation. The decrease of the emission time with increasing mass 
asymmetry was observed also for 3-body events in the symmetric reactions 
$^{100}$Mo+$^{100}$Mo and $^{120}$Sn+$^{120}$Sn around 20 AMeV \cite{Casini}, 
where, on the other hand, the typical spatial configuration with the small 
fragment located between the two large fragments suggests the occurence 
of dynamical effects specific to emission from the symmetric binary 
configuration. 

At excitation energies above the multifragmentation threshold, 
the SMM scenario leads to competition of the hot multifragment partitions  
with the single hot thermally equilibrated fragment ( practically identical 
to the compound nucleus in traditional sense ).   
Within the sequential binary decay scenario with finite range 
fragment emission barriers, as implemented in GEMINI, the onset 
of multifragment channels is much smoother. Apart from the reduction
of the fission barrier, the angular momentum of the hot residue 
leads to a reduction of the thermal excitation energy due to 
the energy of rotational motion. However, the values of angular momentum 
in Fig. \ref{bdep}b imply that in the present work the rotational energy 
typically does not exceed 5 MeV and thus does not influence the 
results, especially at mid-central to central collisions which 
lead to production of heavy residues with A$<$90. 
An interesting aspect of the excitation energy versus mass distributions, 
shown in Fig. \ref{excimp}, is the gradual change of the governing pattern 
from the essentially Wilczynski plots at peripheral 
impact parameters to the pattern typical for incomplete fusion 
scenario at central impact parameters. Thus, the success of SMM implies 
that the residues with A$<$90 observed at forward angles originate 
from incomplete fusion. 

For further comparison, the calculations using only the model 
of deep inelastic transfer combined with both GEMINI and SMM, 
analogous to DIT/GEMI\-NI simulation used in our recent work \cite{GSKrNi}, 
have been carried out. In the case of the nearly symmetric reaction 
$^{86}$Kr+$^{64}$Ni 
at 25 AMeV \cite{GSKrNi}, the DIT/GEMINI simulations proved to be rather 
successful. Similar success was achieved in the reaction 
of 20 AMeV $^{124}$Sn + $^{124}$Sn \cite{gsemis}, especially 
for the residues with A$<$90 on which this work focuses.  
It was verified using both DIT/GEMINI and DIT/SMM that, 
in the reaction $^{124}$Sn + $^{27}$Al at 20 AMeV, 
the yields of residues with A$<$90 can not be reproduced by 
taking into account peripheral processes only, independent of 
which de-excitation code was used. For instance, in the DIT/GEMINI simulations, 
the total unfiltered 
yields are roughly analogous to the simulation presented in 
Fig. \ref{ygem} and the filtered yields again do not reproduce 
the observed yields of residues with A$<$90. In this case, the filtered 
distribution 
exhibits similar decrease in the A$<$90 region but extends further than 
in DIT+ICF/GEMINI case, due to higher excitation energy 
and different angular and momentum distribution of the generated hot nuclei. 
The difference can be attributed mostly to the absence of cooling 
via pre-equilibrium emission, since the products of violent collisions 
are not included in the DIT/GEMINI simulation and do not pass the 
filter procedure in the latter case. On the other hand, the DIT/SMM simulation, 
when compared to DIT+ICF/SMM simulation shown in Fig. \ref{ysmm},  
significantly underpredicts the yields of the residues with A$<$90 
( both unfiltered and filtered ), due to the fact that for peripheral 
collisions, the excitation energy of projectile-like fragments exceeds 
the value of 2 AMeV, where the multifragmentation threshold can be anticipated, 
only very rarely, and due to orbiting, 
such sources move typically at angles away from zero. 

\begin{figure}[p]                    

\includegraphics[width=10.cm,height=10.cm]{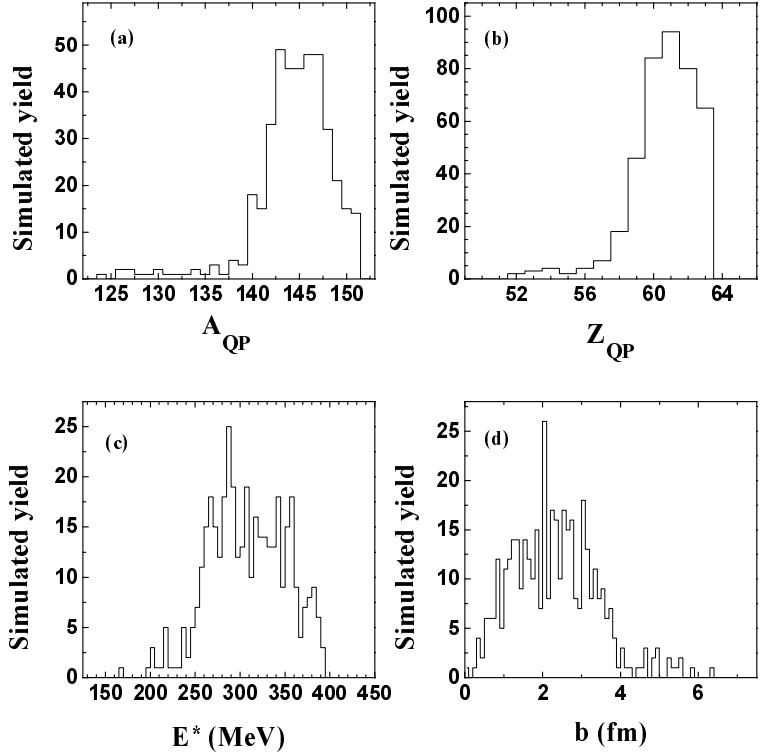}
\centering
\vspace{0.0cm}

\caption{ Mass, charge, excitation energy distributions of hot heavy sources 
contributing to the filtered yield of residues with A$=$65-75 and the 
distribution of contributing impact parameters, as determined by 
backtracing the DIT+ICF/SMM simulation. 
   }

\label{bktrc}

\end{figure}

\begin{figure}[p]                    

\includegraphics[width=10.cm,height=14.cm]{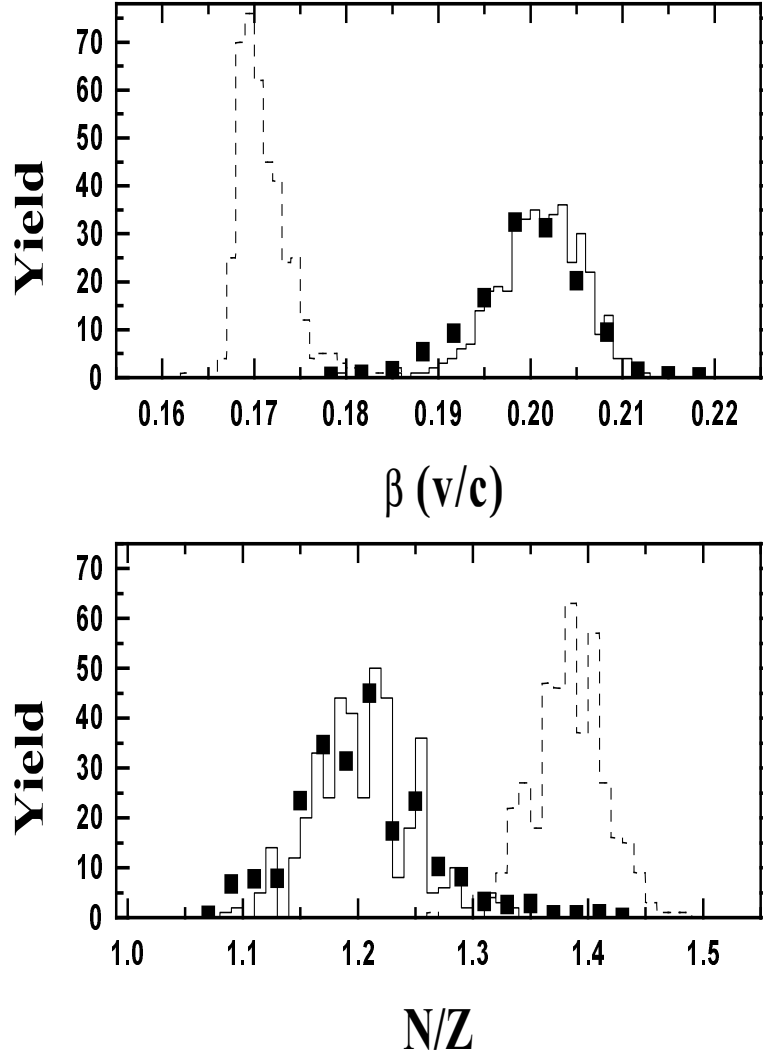}
\centering
\vspace{0.0cm}

\caption{ 
The observed ( symbols ) and filtered ( solid histograms ) 
velocity and N/Z-distributions of residues with A$=$65-75 along with 
the velocity and N/Z-distributions of contributing hot heavy sources 
( dashed histograms ), as determined by backtracing the DIT+ICF/SMM 
simulation. The observed distributions are normalized to the filtered 
distributions.
   }

\label{bktrc2}

\end{figure}

Furthermore, a backtracing procedure was used for the DIT+ICF/SMM 
simulation. In Fig. \ref{bktrc}a,b,c the mass, charge, excitation 
energy distributions of hot heavy sources contributing to filtered 
yield of residues with A$=$65-75 are shown. Fig. \ref{bktrc}d shows the 
distribution of contributing impact parameters. On average, a typical 
contributing hot source can be characterized as a nucleus $^{144}$Nd 
with excitation energy about 310 MeV ( 2.2 AMeV ). The average 
impact parameter is about 2 fm ( $l$ = 44 ) and the average spin 
of the hot source is $J$=30. Thus the results of the backtracing procedure 
are in good agreement with the results of the analysis presented above. 
The observed ( symbols ) and filtered ( solid histograms ) 
velocity and N/Z-distributions of residues with A$=$65-75 along with 
the velocity and N/Z-distributions of contributing hot heavy sources 
( dashed histograms ) are compared in Fig. \ref{bktrc2}. The simulated velocity 
and N/Z-distributions 
reproduce in high detail the experimental ones, thus further demonstrating 
the appropriate description of reaction mechanism within the DIT+ICF/SMM 
simulation and the appropriate treatment of the spectrometer in the 
filtering procedure. The comparison of the observed and filtered velocity 
distribution with the backtraced velocity distributions shows that indeed 
the observed residues correspond to the forward kinematic solution, 
selected by the spectrometer angular acceptance. The momentum acceptance 
of the measurement does not appear to play a role in the filtering procedure. 
The width of the backtraced velocity distribution is rather small and  
corresponds to the incomplete fusion scenario. 
The comparison of the observed and filtered N/Z-distribution with the 
backtraced N/Z-distributions shows a strong shift of the N/Z-ratio towards 
the $\beta$-stability line which is typical for hot heavy sources where 
N/Z-equilibration is achieved during de-excitation. The widths 
of the distributions differ only slightly and the narrow N/Z-distributions 
correspond to the incomplete fusion scenario. 

Since the experimental effect is determined by residues originating 
from violent collisions it is interesting to assess to what extent 
the description of incomplete/complete fusion  can be considered adequate. 
As already mentioned above, the ICF calculation \cite{martin2} proved 
rather successful in quantitative description of the available data 
on properties of hot sources created in the mid-central to central collisions. 
Thus, we believe that the description of the entrance channel does not 
bring uncertainties influencing the conclusions of the present study and that 
the experimental data provide information about the  details of 
the de-excitation stage, allowing to distinguish between different scenarios 
based on quantitative observables, such as the yields of heavy residues. 

\begin{figure}[p]                    

\includegraphics[width=10.cm,height=14.cm]{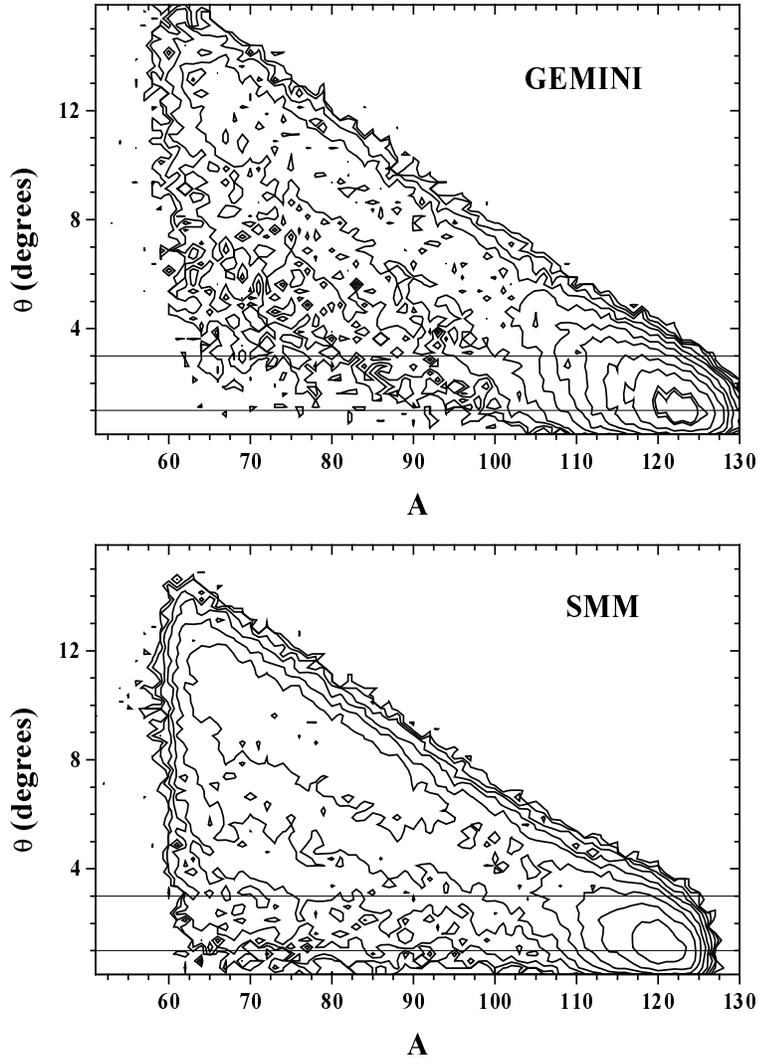}
\centering
\vspace{0.0cm}

\caption{ 
Angle vs mass contour plots of heavy residues 
originating from decay of the hot nucleus $^{144}$Nd 
with excitation energy 310 MeV flying along the beam direction with 
velocity $\beta$=0.17 de-excited 
by GEMINI ( upper panel ) and SMM \hbox{( lower} \hbox{panel ).} 
Successive contours correspond to a decrease  of the yield by a factor of 2.
The two horizontal lines mark the angular acceptance of the MARS separator.    
   }

\label{thcnt}

\end{figure}

\begin{figure}[p]                    

\includegraphics[width=10.cm,height=14.cm]{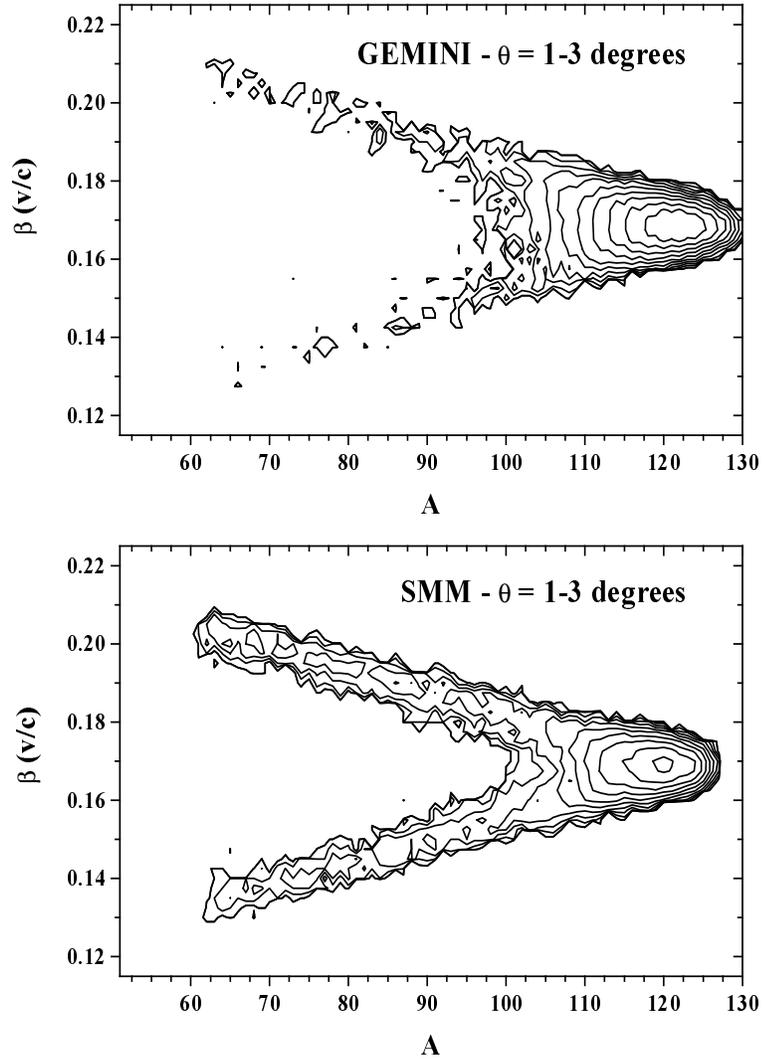}
\centering
\vspace{0.0cm}

\caption{ 
Velocity vs mass contour plots for the simulated 
residues with $\theta$=1-3 deg originating from decay 
of the hot nucleus $^{144}$Nd with excitation energy 310 MeV 
flying along the beam direction with velocity $\beta$=0.17 de-excited 
by GEMINI ( upper \hbox{panel )} and SMM ( lower panel ). 
Successive contours correspond to a decrease  of the yield by a factor of 2.
   }

\label{vcnt}

\end{figure}

\begin{figure}[p]                    

\includegraphics[width=10.cm,height=14.cm]{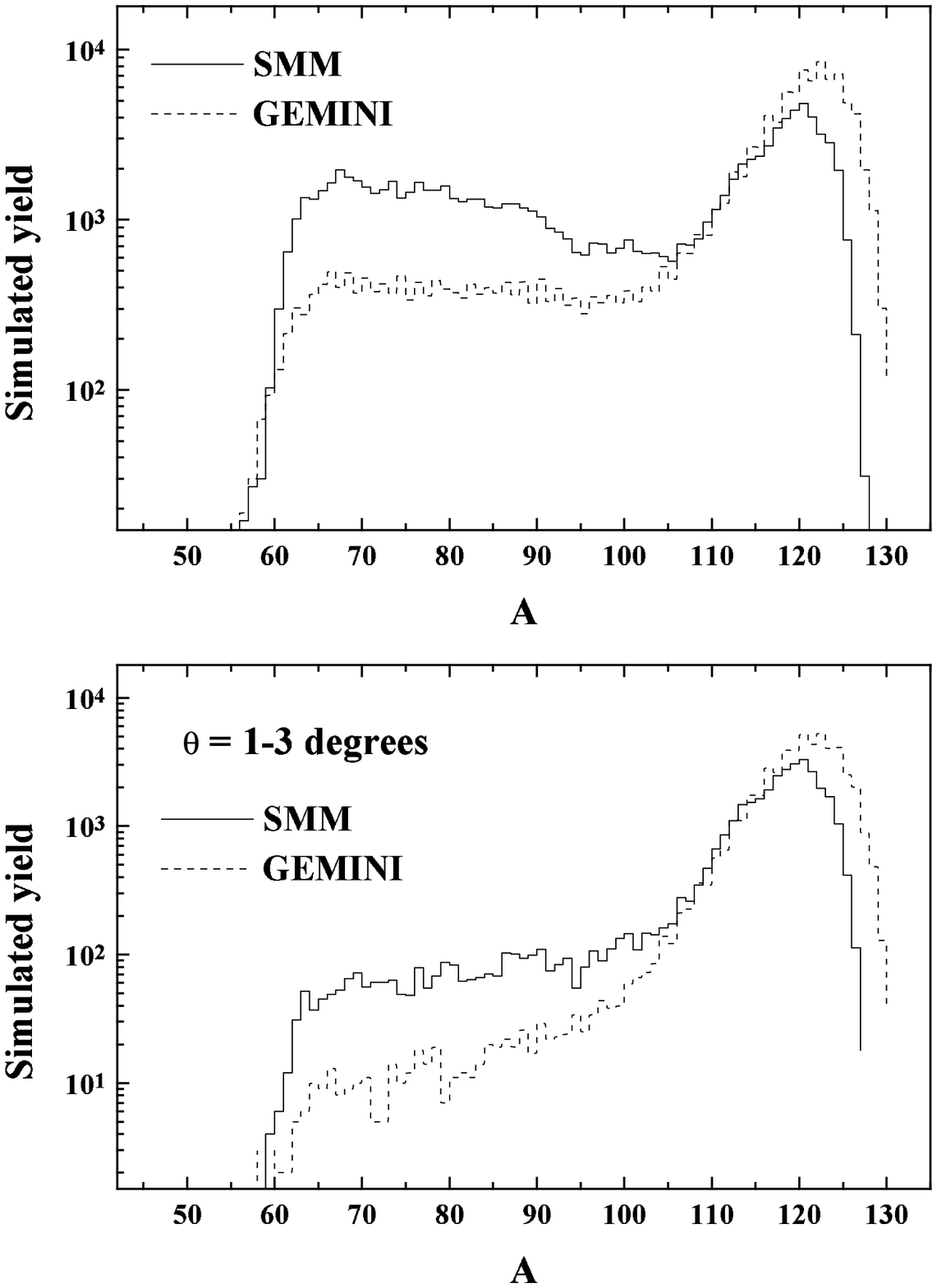}
\centering
\vspace{0.0cm}

\caption{ 
Unfiltered yields and yields filtered by MARS angular acceptance 
( upper and lower panels respectively ) of heavy residues 
originating from the decay of the hot nucleus $^{144}$Nd 
with excitation energy 310 MeV flying along the beam direction with 
velocity $\beta$=0.17 de-excited 
by GEMINI ( dashed line ) and SMM ( solid \hbox{line ).} 
   }

\label{ycnt}

\end{figure}

In order to illustrate in detail the difference in the de-excitation scenarios 
of sequential binary decay and statistical multifragmentation, 
we carried out simplified calculations. 
In agreement with the results of backtracing 
procedure, the hot nucleus $^{144}$Nd with excitation energy 310 MeV 
flying along the beam direction with velocity $\beta$=0.17 was de-excited 
by both GEMINI and SMM. The spin of the hot nucleus was chosen $J$=30, 
in accord with backtracing procedure. The resulting angle versus mass contour 
plots are presented in Fig. \ref{thcnt} in a fashion analogous 
to Fig. \ref{thgemsmm}. The behavior observed in Fig. \ref{thgemsmm} 
is essentially reproduced. Again, the SMM leads to larger yields 
of residues with A$<$90 at angles corresponding to the MARS acceptance. 
The effect of MARS angular acceptance is demonstrated in Fig. \ref{vcnt} 
where the velocity versus mass contour plots are shown for the simulated 
residues with $\theta$=1-3 deg acceptance. Again the results for the heavy 
residues corresponding to the forward kinematic solution 
( upper arm in Fig. \ref{vcnt} ) are analogous 
to the results of full simulations as shown in Figs. \ref{vgem},\ref{vsmm}.  
The heavy residues corresponding to the backward kinematic solution 
( lower arm ) were not detected in the experiment and thus are rejected 
by the full MARS filtering procedure for the current experiment. 
As suggested above, the selection caused by the angular acceptance of MARS 
indeed seems to be the most crucial criterion of the filtering procedure 
for the heavy residues investigated in the present work. 
In Fig. \ref{ycnt} both unfiltered yields 
and yields filtered by the MARS angular acceptance ( upper and lower panels, 
respectively ) for GEMINI and SMM simulations are shown. As one can see, 
the unfiltered yield of heavy residues with A$<$90 is higher in the SMM 
simulation and this is essentially preserved after filtering by MARS angular 
acceptance. The essential difference between the results of SMM and GEMINI in 
the simplified calculation appears to be in the production rate of heavy 
residues with A$<$90, in good agreement with the conclusions set 
in the discussion of Fig. \ref{prob}. Using a backtracing procedure 
of this simplified SMM simulation it was determined that the most probable 
hot fragment partition contributing to the filtered yield is essentially a 
ternary configuration resembling the ternary fission with two heavier 
fragments accompanied by a lighter third particle. The same conclusion 
applies to backtracing of the full simulation. In general, the simplified 
and full simulations lead essentially to the same results, which is even 
more convincing when taking into account that the source characteristics 
have been chosen using backtracing of SMM simulation but nevertheless 
the result is equivalent also for the simulations using GEMINI. 
Based on the simplified simulation, we assure that the reason for different 
production rates derives from the difference of physical scenarios 
used by the de-excitation codes. 

In the particular case of the GEMINI code it is of importance 
to establish to what extent an adjustment of the parameter set could be 
justified in order to improve the agreement of the simulation with experiment. 
As already illustrated in the discussion of 
Fig. \ref{prob}, the parameter essentially determining the decay 
rate of channels with heavy residues with A$<$90 is the fission 
barrier height. The comparison with the recent high-quality experimental 
data on asymmetric fission barriers \cite{BarMor} suggests that the finite 
range fission barriers used in the GEMINI code are in reasonable 
agreement with experiment. The asymmetry dependence of the experimental 
barriers is reproduced well, while the experimental values are typically  
higher by 1-2 MeV. Thus, there is no physical justification for an enhancement 
of the decay rate by lowering the fission barriers. In a similar fashion, 
the choice of the level density parameter of Lestone \cite{Lestone} 
is in agreement with the systematics of effective level density 
parameters obtained using the experimental data on particle emission 
spectra \cite{Lestone,ShNat}. 
The value of the level density parameter in the saddle point configuration 
is typically larger ( by 3-4 \% ) than in the particle evaporation channels, 
due to the temperature dependence of the level density parameter of Lestone. 
Such a calculation is essentially equivalent to the calculation 
with temperature indepedent level density parameter and a$_{f}$/a$_{n}$ 
ratio according to models taking into account the surface area increase 
in the saddle configuration \cite{AfAnMod}. However, several experimental  
works \cite{AfAn} suggest that a value of a$_{f}$/a$_{n}$=1 is good 
approximation at excitation energies in the saddle configuration above 40 MeV, 
mostly based on statistical model analysis of evaporation residue cross 
sections and pre-fission neutron multiplicity. In any case, 
the further increase of a$_{f}$/a$_{n}$ ratio 
does not appear to be justified. 
Thus, the parameter set used can be considered appropriate 
as demonstrated in other works \cite{GSKrNi,gsemis} where heavy residues 
originating from peripheral collisions have been investigated. 
In these deep-inelastic collisions, the details of de-excitation 
are overshadowed by the broad excitation energy, angle 
and velocity distributions. 

The insight into the difference between the sequential binary and 
simultaneous decay can be obtained using the conclusions of the 
paper of Lopez and Randrup \cite{LopRand}, where a statistical model 
of sequential binary decay is compared to a more general statistical 
model where a transitional state formalism is extended to 
the multifragment partition with any number of fragments. 
The authors observe that the transitional 
state model of multifragmentation leads also to a more general expression 
for the decay rate of binary channels leading to additional enhancement 
at high temperatures due to additional available degrees of freedom 
when compared to the binary decay formula used in the GEMINI code. 
In a similar fashion we 
can conclude that the decay rate of channels with higher fragment 
multiplicities will be enhanced in the more general model of Lopez and Randrup 
when compared to the corresponding de-excitation chain 
with binary splits, due to still higher number of available degrees of freedom. 
When considering, as suggested by the authors, that the transitional state 
model of multifragmentation presented in \cite{LopRand} is an alternative 
formulation to the models of nuclear disassembly such as statistical 
multifragmentation model \cite{smm}, on which the SMM code is based, 
we can attribute the success of the SMM simulation in describing 
the experimental yields to the proper treatment of the available 
degrees of freedom in de-excitation of hot sources with the highest 
excitation energies created in the most violent collisions. 

From the comparison of the measured yields with the results of simulations, 
one can conclude that the heavy residues with A$<$90 observed in the 
experiment are produced in violent collisions at low impact parameters, 
where a single hot source is created by incomplete fusion which further 
de-excites via simultaneous multifragmentation. The production of such 
residues can not be explained either by peripheral collisions or by 
fission and sequential decay of the hot source. The DIT+ICF/SMM simulation 
is shown to be a realistic model framework for the description of 
asymmetric nucleus-nucleus collisions at projectile energies around 
the Fermi energy, offering a quantitative description of the experimental data.

 \section{Summary and conclusions}

In the present work, the cross sections and velocity distributions 
of heavy residues from the reaction of 20 AMeV $^{124}$Sn on $^{27}$Al 
have been measured at forward angles using the MARS recoil separator 
at Texas A\&M. A consistent overall description 
of the measured cross sections and velocity distributions was achieved 
using a model calculation employing the concept 
of deep-inelastic transfer for the primary stage of the peripheral 
collisions, pre-equilibrium emission and incomplete fusion for the primary 
stage of the more violent central collisions and the  statistical 
model of multifragmentation for the deexcitation stage. 
An alternative calculation employing the model of 
the sequential binary decay could not reproduce the observed yields of the 
residues from violent collisions (A$<$90) since the angular distribution 
of the generated residues does not extend far enough toward zero angle. 
The success of the statistical multifragmentation model can be attributed 
to a better treatment of decay rates and kinematic properties of 
de-excitation channels, leading to heavy residues with A$<$90, at excitation 
energies around the multifragmentation threshold, within the scenario 
where various multifragment partitions compete according to their 
statistical decay widths. The decay widths for sequential binary decay 
exhibit restrictions inherent to the physical picture of the asymmetric binary 
fission. As demonstrated in the present work, a high resolution measurement 
using a kinematic separator is a sensitive method allowing to investigate the 
details of de-excitation of hot nuclei created in violent collisions.


\section{Acknowledgment}


We gratefully 
acknowledge the support of the  operations staff of the Cyclotron Institute
during the measurements. We express our gratitude to L. Tassan-Got 
for the opportunity to use his DIT code, to R. Charity for the use 
of the GEMINI code and to A.S. Botvina for the use of the SMM code. 
Financial support for this work was given, in part, by the U.S. Department 
of Energy under Grant No. DE-FG03-93ER40773, by the Robert A. Welch 
Foundation under Grant No. A-1266 and by Slovak Scientific Grant Agency 
under Grant No. VEGA-2/1132/21. M.V. would like to express his gratitude 
to the Cyclotron Institute of Texas A\&M University for hospitality 
during his stay.  



\newpage

\begin{thebibliography}{00}


\bibitem{bazin}
D. Bazin {\it et al.}, 
Nucl. Phys. {\bf A515} (1990) 349.
\bibitem{faure}
B. Faure-Ramstein {\it et al.}, 
Nucl. Phys.  {\bf A586} (1995) 533.
\bibitem{pfaff} 
R. Pfaff {\it et al.}, 
Phys. Rev. {\bf C53} (1996) 1753.
\bibitem{karl}
K.A. Hanold {\it et al.},  Phys. Rev.  {\bf C52} (1995) 1462.

\bibitem{george}
G.A. Souliotis {\it et al.}, 
Phys. Rev. {\bf C57} (1998) 3129.

\bibitem{george2}
G.A. Souliotis {\it et al.}, 
Nucl. Phys. {\bf A705} (2002) 279.

\bibitem{george3}
G.A. Souliotis, Physica Scripta {\bf T88} (2000) 153.

\bibitem{spectra}                                                    
K. Aleklett {\it et al.}, 
Nucl. Phys.  {\bf A499} (1989) 591.
\bibitem{sarar}
W. Loveland {\it et al.}, 
Phys. Rev. {\bf C41} (1990) 973.

\bibitem{GSKrNi}
G.A. Souliotis {\it et al.}, Phys. Lett. {\bf B543} (2002) 163. 

\bibitem{granier}
O. Granier {\it et al.}, Nucl. Phys. {\bf A481} (1988) 109.
\bibitem{barz}
H. Barz {\it et al.}, 
Phys. Rev. {\bf C46} (1992) R42.
\bibitem{yokoyama}
A. Yokoyama {\it et al.}, 
Phys. Rev. {\bf C46} (1992) 647.
\bibitem{marchetti}
A.A. Marchetti {\it et al.}, Phys. Rev. {\bf C48} (1993) 266.
\bibitem{aboufirassi}
M. Aboufirassi {\it et al.}, LPCC 93-14, September, 1993; 
J.F. LeColley {\it et al.}, Phys. Lett. {\bf B325} (1994) 317.
\bibitem{garcia} 
E.J. Garcia-Solis {\it et al.}, 
Phys. Rev. {\bf C52} (1995) 3114.
\bibitem{baldwin}
S.P. Baldwin {\it et al.}, Phys. Rev. Lett. {\bf 74} (1995) 1299.
\bibitem{morjean}
M. Morjean {\it et al.}, Nucl. Phys.  {\bf A591} (1995) 371.
\bibitem{enterria}
D.G. d'Enterria {\it et al.}, 
Phys. Rev. {\bf C52} (1995) 3179.

\bibitem{MVSiSn}
M. Veselsky {\it et al.}, Phys. Rev. {\bf C62} (2000) 64613.

\bibitem{MARS}   R.E. Tribble, R.H. Burch and C.A. Gagliardi,
                 Nucl. Instr. and  Meth. {\bf A285} (1989) 441; 
		 R.E. Tribble, C.A. Gagliardi and W. Liu, 
                 Nucl. Instr. and  Meth. {\bf B56/57} (1991) 956.

\bibitem{Wilcke} 
W.W. Wilcke {\it et al.},
                 At. Data and Nucl. Data Tables {\bf 25} (1980) 389.

\bibitem{Greg}  G. Chubarian, private communication.
            
\bibitem{Hubert}
F. Hubert, R. Bimbot and H. Gauvin, 
At. Data and Nucl. Data Tables  {\bf 46} (1990) 1; 
Nucl. Instrum. Methods  {\bf B36} (1989) 357.
\bibitem{Leon} 
A. Leon {\it et al.},  
At. Data and Nucl. Data Tables {\bf 69}  (1998) 217.


\bibitem{GSfission}
 G.A. Souliotis {\it et al.},  Phys. Rev.  {\bf C55}  (1997) R2146.

\bibitem{martin2}
 M. Veselsky, Nucl. Phys. {\bf A705} (2002) 193. 

\bibitem{tassan}
 L. Tassan-Got and C. Stefan,  Nucl. Phys. {\bf A524} (1991) 121.

\bibitem{GEMINI}
 R. Charity {\it et al.}, Nucl. Phys.  {\bf A483}  (1988) 391.
 The version of GEMINI included modifications made up to July, 1998.
\bibitem{Moretto}
 L.G. Moretto, Nucl. Phys. {\bf A247} (1975) 211.

\bibitem{smm}
J.P. Bondorf {\it et al.}, Phys. Rep. {\bf 257} (1995) 133. 

\bibitem{Lestone}
 J. Lestone, Phys. Rev.  {\bf C52}  (1995) 118.

\bibitem{Marmier}
 P. Marmier and E. Sheldon, Physics of Nuclei and Particles, 
 Volume I (Academic, New York, 1970) p. 15.

\bibitem{SierkAsym} 
A.J. Sierk, Phys. Rev. Lett. {\bf 55} (1985) 582.

\bibitem{sierk}
A.J. Sierk, Phys. Rev. {\bf C33} (1986) 2039.

\bibitem{barash}
V.S. Barashenkov, A.S. Iljinov, V.D. Toneev, F.G. Gereghi, 
Nucl. Phys. {\bf A206} (1973) 131.

\bibitem{Adeev}
G.D. Adeev {\it et al.}, Preprint INR 816/93, Moscow, 1993.

\bibitem{Gui}
M. Gui {\it et al.}, Phys.Rev. {\bf C48} (1993) 1791. 

\bibitem{Casini}
G. Casini {\it et al.}, Phys.Rev. Lett. {\bf 71} (1993) 2567. 

\bibitem{gsemis}
G.A. Souliotis {\it et al.}, Nucl. Instr. and  Meth. {\bf B204} (2003) 166.

\bibitem{BarMor}
K.X. Jing  {\it et al.}, Nucl. Phys. {\bf A645} (1999) 203;
T.S. Fan  {\it et al.}, Nucl. Phys. {\bf A679} (2000) 121.

\bibitem{ShNat}
S. Shlomo and J.B. Natowitz, Phys. Lett. {\bf B252} (1990) 1987; 
S. Shlomo and J.B. Natowitz, Phys. Rev. {\bf C44} (1991) 2878. 

\bibitem{AfAnMod} 
A.V. Ignatyuk, M.G. Itkis, V.N. Okolovich, G.N. Smirenkin,and A.S. Tishin, 
Yad. Fiz. 21 (1975) 1185 (Sov. J. Nucl. Phys. 21 (1975) 612); 
J. Toke and W.J. Swiatecki, Nucl. Phys. A372, 141 (1981).

\bibitem{AfAn}
A.N. Andreyev {\it et al.}, Nucl. Phys. {\bf A620} (1997) 229; 
D.D. Bogdanov {\it et al.}, Phys. of At. Nucl. {\bf 62} (1999) 1794; 
M. Veselsky {\it et al.}, AIP Conf. Proc. 447 (1998) 291.

\bibitem{LopRand}
J.A. Lopez and J. Randrup, Nucl. Phys. {\bf A503} (1989) 183.

\end{thebibliography}
\end{document}